\documentclass[twocolumn,pra,showpacs,superscriptaddress,amsmath,amssymb]{revtex4}

\usepackage{graphicx}% Include figure files
\usepackage{color}

\newcommand{\Qd}{Q}
\newcommand{\Pd}{P}
\newcommand{\qs}{q}
\newcommand{\ps}{p}

\begin{document}

\title{Weak values in a classical theory with an epistemic restriction}% Force line breaks with \\

\author{Angela Karanjai}%
\author{Eric G. Cavalcanti}%
\author{Stephen D. Bartlett}%
\affiliation{Centre for Engineered Quantum Systems, School of Physics, The University of Sydney, Sydney, NSW 2006, Australia}%
\author{Terry Rudolph}%
\affiliation{Department of Physics, Imperial College London, London SW7 2AZ, United Kingdom}%

\date{15 July 2015}

\begin{abstract}
Weak measurement of a quantum system followed by postselection based on a subsequent strong measurement gives rise to a quantity called the \emph{weak value}: a complex number for which the interpretation has long been debated. We analyse the procedure of weak measurement and postselection, and the interpretation of the associated weak value, using a theory of classical mechanics supplemented by an epistemic restriction that is known to be operationally equivalent to a subtheory of quantum mechanics. Both the real and imaginary components of the weak value appear as phase space displacements in the postselected expectation values of the measurement device's position and momentum distributions, and we recover the same displacements as in the quantum case by studying the corresponding evolution in our theory of classical mechanics with an epistemic restriction.  By using this epistemically restricted theory, we gain insight into the appearance of the weak value as a result of the statistical effects of post selection, and this provides us with an operational interpretation of the weak value, both its real and imaginary parts.  We find that the imaginary part of the weak value is a measure of how much postselection biases the mean phase space distribution for a given amount of measurement disturbance.  All such biases proportional to the imaginary part of the weak value vanish in the limit where disturbance due to measurement goes to zero.  Our analysis also offers intuitive insight into how measurement disturbance can be minimised and the limits of weak measurement. 
\end{abstract}

\pacs{Valid PACS appear here}% PACS, the Physics and Astronomy
\maketitle

\section{\label{sec:level1}Introduction}

One of the most distinctive features of quantum mechanics is the necessary disturbance to the quantum state associated with any measurement that acquires information about the state.  This information gain-disturbance relation places restrictions on what types of measurements are allowed within quantum theory.  Weak measurements are a limiting case of a class of measurements with which it is possible to measure the average value of some observable using an ensemble of particles, all prepared in the same initial state, with minimal disturbance to the state of each individual particle. Such measurements have a long history in quantum theory (see, for example, Refs.~\cite{weak1, weak2, weak3, weak4}).  

Performing a weak measurement leaves the state of the particle largely undisturbed, and one can consider performing a subsequent measurement, possibly of a different observable.  Consider an ensemble of particles prepared in the same state $|\Psi\rangle$ subjected to a weak measurement of observable $A$ followed by a projective measurement of observable $B$, and then postselecting only those experiments corresponding to a specific outcome $|b\rangle$ of $B$.  

It is within the context of such experiments that Aharonov \emph{et al.}~\cite{aharonov} introduced the \emph{weak value}, 
\begin{equation}
\label{eq:1a}
\langle\hat{A}\rangle _W = \frac{\langle b|\hat{A}|\Psi\rangle}{\langle b|\Psi\rangle}\,,
\end{equation}
as the measurement outcome of the observable $A$ for the preselected and postselected ensemble.  Subsequently, there was considerable debate over the meaning of this weak value, as it is in general a complex number.  An operational interpretation of the weak value as a complex number, whose real and imaginary parts manifest as shifts in the average position and momentum of the post selected measurement devices, was given by Jozsa~\cite{complex}.  The interpretation of the real part of the weak value as the conditional expectation value of the variable $A$~\cite{meaning, time, average, interval, reply,Dressel} has been used in analysing counterfactual quantum paradoxes~\cite{paradox1, paradox2, paradox3, cat}. The imaginary part of the weak value has been connected to the shift in momentum of the pointer associated with measurement disturbance~\cite{imaginary1, imaginary2}.  Another debated property of the weak value is that it is not constrained by the eigenvalue spectrum of the variable, that is, the weak value can be larger than the largest eigenvalue of the variable~\cite{big}.  Such anomalous weak values have been considered for signal amplification~\cite{amplification1, amplification2, noamplification1, noamplification2}.  The appearance of anomalous weak values can be used to provide a proof of contexuality~\cite{anomolous}, which suggests that interpreting the real part of the weak value as a conditional expectation value needs to be reevaluated.

Much of the difficulty in interpreting the weak value may be because it seeks to analyse the measurement outcomes of two noncommuting observables on a given state of a particle, which is known to be problematic in quantum theory due to the lack of an ontology for measurement outcomes associated with observables.  It is worthwhile, then, to consider whether the weak value can arise in a theory that does possess a clear ontology.  Recently, it has been shown that similar features to the weak value in quantum theory can arise within a simple statistical model supplemented with a backaction due to measurement~\cite{classical2}, suggesting that the weak value is a statistical feature in theories involving measurement disturbance.  The suggestion that weak values can arise in a classical analog is controversial~\cite{Dressel,classical2resp,Vaidman}, and it has been argued that weak values have no analog in classical statistics~\cite{Vaidman}.

In this paper, we analyse weak values using a theory of classical mechanics (thereby possessing a clear ontology) supplemented with a restriction on the observer's knowledge.  This theory is the \emph{epistemically restricted Liouville (ERL)} mechanics of Ref.~\cite{erl}, and it is known to reproduce many of the features of quantum measurement.  In this theory, all particles evolve under classical equations of motion and it is operationally equivalent to gaussian quantum mechanics; this connection is best seen through the description of gaussian quantum mechanics using nonnegative Wigner functions.  Notably, the epistemic restriction provides a sensible notion of weak measurement within the ERL theory, one that directly reproduces many of the key features of quantum weak measurement.  We emphasise that ERL theory adds neither extra stochasticity to classical dynamics nor any additional disturbance mechanism; rather, all of the features analogous to quantum theory appear naturally within a deterministic theory supplemented only by ignorance on the part of the observer.

Within ERL theory, as in quantum mechanics, we find that the weak value appears operationally as shifts in the mean position and momentum distributions of the measurement device upon postselection (as first discussed by Josza~\cite{complex}).  The analysis in the ERL theory gives us a direct interpretation of the origin of these shifts, and thus of the weak value.  Specifically, the real component of the weak value represents the shift in the position of the measurement device as a result of its interaction with the measured particle, as expected from a measurement.  The imaginary component of the weak value, however, quantifies not the result of any dynamical changes to the measurement device but simply a bias on the distribution of the measurement device as a result of postselection.  That is, we have an operational interpretation of the imaginary part of the weak value as a measure of how much postselection will bias the distribution of the measurement device.  The weak value is not a unique feature of quantum theory, but can arise in other theories that possess a restriction or limitation on the observer's knowledge of the initial state of the particle or the measurement device, which is arguably a very natural physical restriction.

We note that anomalous weak values do not appear in our analysis, as all observables in our model possess an unbounded spectrum.  Consistent with the results of Ref.~\cite{anomolous}, our model is also noncontextual: the ERL mechanics provides an explicit noncontextual ontological model for all procedures described here.

\section{Weak measurements and the weak value}
\label{sec:level2}

In this section, we introduce the formalism of weak measurements within quantum theory, as well as briefly introduce the weak value.  We first review the standard formalism for von Neumann measurements, including strong (projective) measurements, and then introduce weak measurements within this model.  We then demonstrate the appearance of the weak value (both real and imaginary parts) in the conditional expectation values of the position and momentum of the measurement device after postselection.

\subsection{The Von Neumann measurement model}
\label{sec:level2a}

Here, we review the framework of quantum measurement, wherein an observable is coupled to a measurement device followed by a projective measurement of the measurement device's position.  With this framework, we can describe both strong (projective) as well as arbitrarily weak measurements.  

We describe the measurement device by a one-dimensional quantum system with canonical position observable $\hat{\Qd}$ and momentum observable $\hat{\Pd}$ satisfying $[\hat{\Qd},\hat{\Pd}]=i\hbar$.  In what follows, we choose units such that $\hbar = 1$.  We denote position eigenstates by $|\Qd\rangle$ satisfying $\hat{\Qd}|\Qd\rangle = \Qd|\Qd\rangle$.  

Consider the initial state of the wavefunction of the measurement device to be a pure gaussian state with mean position 0. The most general form of such a state is
\begin{equation} 
  \label{eq:1a} 
  | \Phi\rangle=\int \phi(\Qd)|\Qd \rangle\,d\Qd \,, 
\end{equation} 
where 
\begin{equation} 
  \phi(\Qd)\propto \exp\Bigl(\frac{(i \Omega -1)\Qd^2}{4\Delta_{\Qd}^2}+i\Qd \mu_{\Pd}\Bigr),
\end{equation}
up to an irrelevant normalisation constant.  Here, $\Delta_{\Qd}$ is the standard deviation of the position of the device, $\mu_{\Pd}$ is the mean momentum of the measurement device and $\Omega$ is the covariance of the device
\begin{equation} 
  \label{eq:2b} 
  \Omega=\langle\hat{\Pd}\hat{\Qd}+\hat{\Qd}\hat{\Pd}\rangle-2\langle\hat{\Pd}\rangle \langle\hat{\Qd}\rangle\,.\end{equation}
As the measurement device is in a pure gaussian state, it saturates the uncertainty principle and hence the uncertainty in its momentum is
\begin{equation} 
  \label{eq:3c} 
  \Delta_{\Pd}^2=\frac{1+\Omega^2}{4\Delta_{\Qd}^2}.
\end{equation}

Such a device can be used to measure a quantity given by the particle observable $\hat{A}$ by coupling the particle and device via the interaction Hamiltonian
\begin{equation} 
  \label{eq:4d}
  \hat{H}_{\rm int}=\chi(t)\hat{\Pd}\otimes\hat{A}\, ,
\end{equation}
where $\hat{\Pd}$ acts on the measurement device and $\hat{A}$ acts on the particle. 

An impulsive measurement is performed over a time interval $[0,t_0]$ such that
\begin{equation} 
  \label{eq:5e} 
  \int^{t_0}_0\chi(t)\,dt=g,
\end{equation}
where $g$ is the effective interaction strength. Consider an initial state of the particle given as $|\Psi\rangle=\sum_j\alpha_j|a_j\rangle$, where $|a_j\rangle$ is an eigenstate of $\hat{A}$ with eigenvalue $a_j$.  After the interaction, the state of the device and particle will be
\begin{equation}
\label{eq:6f}
e^{-i g\hat{\Pd}\hat{A}}|\Phi\rangle|\Psi\rangle =\sum_j\alpha_j \Bigl( \int\phi(\Qd-g a_j)|\Qd\rangle d\Qd \Bigr) |a_j\rangle.
\end{equation}
Consider the case where the initial uncertainty in the position of the pointer $\Delta_{\Qd}$ is zero, and thus $|\Phi\rangle$ is a position eigenstate with eigenvalue $0$.  In this case, \begin{equation}
\label{eq:7g}
\lim_{\Delta_{\Qd} \to 0} e^{-i g\hat{\Pd}\hat{A}}|\Phi\rangle |\Psi\rangle = \sum_j\alpha_j|\Qd =g a_j\rangle|a_j\rangle. 
\end{equation}
After the interaction, the measurement device's position is maximally entangled with the eigenstates of $\hat{A}$ of the particle. A projective measurement of the position of the device pointer perfectly resolves the eigenvalue of $\hat{A}$, and collapses the state of the particle into an eigenstate of $\hat{A}$.  This measurement, then, corresponds to a strong, projective measurement of $\hat{A}$ on the particle.

\subsection{Weak measurements}
\label{sec:level2aa}

The weak measurement limit is the opposite limit of this strong measurement, and aims to reduce the disturbance to an arbitrarily small amount at the expense of a correspondingly small information gain. Within the measurement model described above, we introduce two different ways in which a measurement can be made weak.  First, each particle can be coupled arbitrarily weakly to a measuring device by using a vanishingly small interaction strength. Alternatively, a weak measurement can also be obtained by using an initial state of the measurement device with $\mu_{P}=0$ and $\Delta_{\Pd}\to 0$, which would imply $\Delta_{\Qd}\to\infty$ from \eqref{eq:3c}.  (While these two limits lead to identical measurement statistics within quantum theory, we explore each of them separately, as they will correspond to different processes in the context of the epistemically-restricted theory of classical mechanics explored in the next section.) In both of these limits, the disturbance caused by measurement, as well as the amount of information gained, are both very small. If this measurement is repeated on a large number of particles, each prepared in the same initial state $|\Psi\rangle$, it is possible to measure the average value of an observable $A$ of the ensemble of particles with arbitrary accuracy as the number of particles becomes large.

\subsection{The quantum weak value}
\label{sec:level2b}

\begin{figure}[t]
\includegraphics[width=0.45\textwidth]{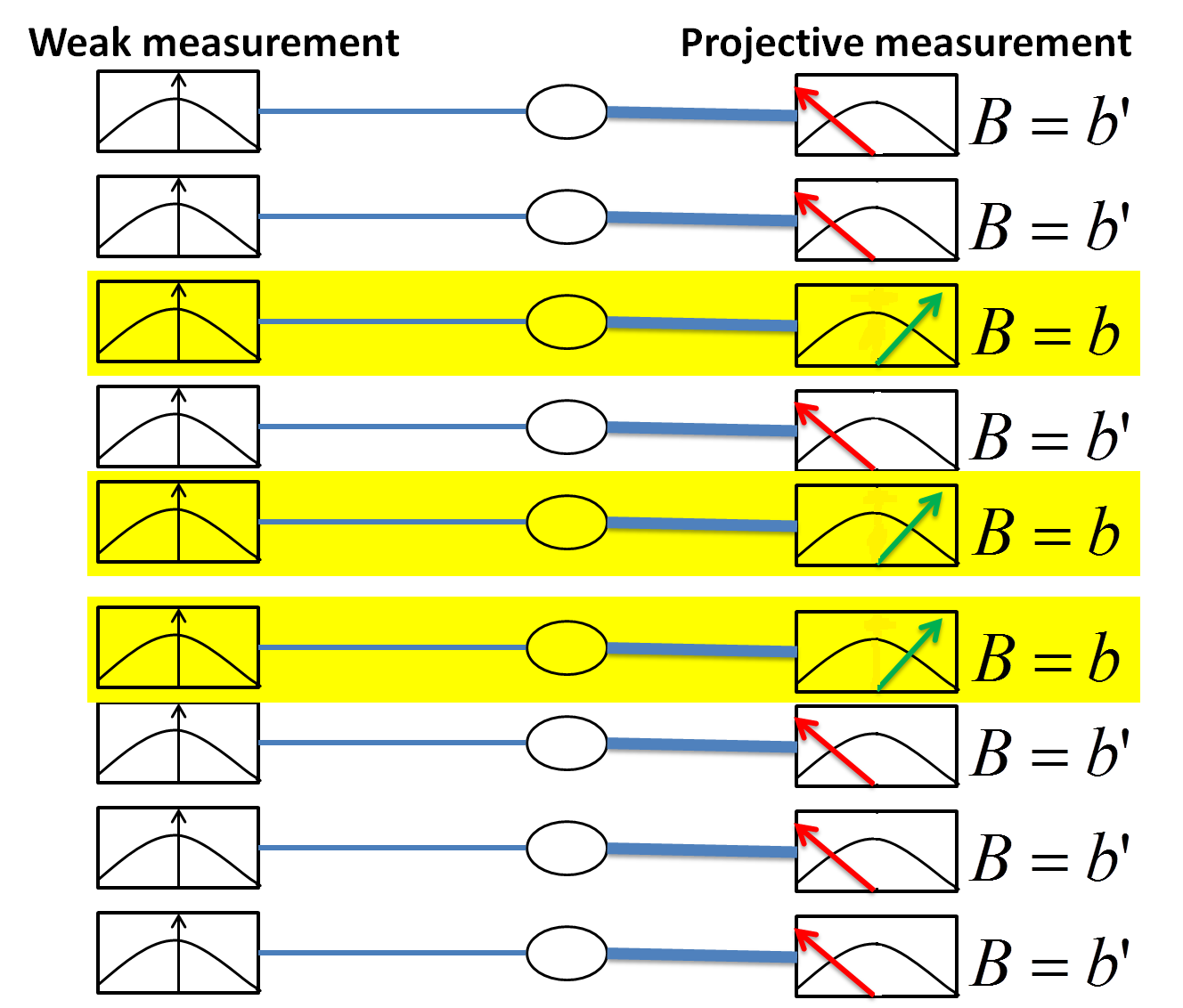}
\caption{An illustration of the operational protocol.  The particles are all prepared in the same initial state. They are first measured weakly by measurement devices shown on the left, followed by a projective measurement of observable $B$ shown on the right. The shift in the mean position of the measurement devices that weakly interacted with the particles on the postselected set (shaded) corresponding to outcome $b$ has terms proportional to the real and imaginary parts of the weak value. }\label{fig1}
\end{figure}

With the concept of weak measurement, we now derive the weak value, with an emphasis on the difference between the two weak measurement methods described above.

The \emph{weak value} arises in a measurement scenario wherein a weak measurement of an observable $A$ is followed by a strong (projective) measurement of another observable $B$, together with postselection on a particular outcome labelled by eigenvalue $b$ of the measurement of $B$. The observable $B$ on which results are postselected need not commute with the variable $A$ weakly measured as illustrated in Fig 1.  In such a situation, the weak value is defined~\cite{aharonov} to be
\begin{equation}
\label{eq:12j}
\langle\hat{A}\rangle_W = \frac{\langle b|\hat{A}|\Psi\rangle}{\langle b|\Psi\rangle}\,,
\end{equation}
where $\hat{B}|b\rangle=b|b\rangle$. Using $|\Psi\rangle = \sum_j \alpha_j |a_j \rangle$, the real part of this complex number is
\begin{equation}
 \text{Re}[\hat{A}_W]=\frac{\sum_{j,l}\alpha_{j}\alpha_{l}^*\langle b|a_{j}\rangle\langle a_{l}|b\rangle (\frac{a_{j}+a_{l}}{2})}{\sum_{j,l}\alpha_{j}\alpha_{l}^*\langle b|a_{j}\rangle\langle a_{l}|b\rangle}\,,
\end{equation}
and its imaginary part is
\begin{equation}
\text{Im}[\hat{A}_W] = \frac{\sum_{j,l}\alpha_{j}\alpha_{l}^*\langle b|a_{j}\rangle\langle a_{l}|b\rangle (\frac{a_{j}-a_{l}}{2i})}{\sum_{j,l}\alpha_{j}\alpha_{l}^*\langle b|a_{j}\rangle\langle a_{l}|b\rangle}\,.
\end{equation}

We will now show explicitly how the real and imaginary parts of the weak value appear as phase-space displacements in the mean position and momentum of the postselected distribution of the measurement devices performing the weak measurements.  The unnormalised state of the post selected particles and the devices after weak measurement is
\begin{equation}
  \label{eq:13k}
  (I\otimes|b\rangle\langle b|)e^{-i g\hat{\Pd}\otimes\hat{A}}|\Phi\rangle|\Psi\rangle \,.
\end{equation}
Recall that $|\Phi\rangle$ is a gaussian state of the form of Eq.~(\ref{eq:1a}).  Having postselected particle-measurement device pairs for which the particle is in the final state $|b\rangle$, the selected devices are described by the unnormalised state $|\Phi'\rangle=\sum_j \alpha_j\langle b|a_j\rangle\int\phi(\Qd-g a_j)|\Qd\rangle d\Qd$. The mean position of this device state after postselection on $b$, denoted $\langle \hat{\Qd}\rangle_b$, is
\begin{widetext}
\begin{align}
\label{eq:14l}
\langle \hat{\Qd}\rangle_b &=\frac{\sum_{j,l}\alpha_j\alpha_l^*\langle b|a_j\rangle\langle a_l|b\rangle \int \Qd \phi(\Qd -g a_j)\phi^*(\Qd -g a_l) d\Qd }{\sum_{j,l}\alpha_j\alpha_l^*\langle b|a_j\rangle\langle a_l|b\rangle\int \phi(\Qd -g a_j)\phi^*(\Qd -g a_l) d\Qd } \nonumber \\
&=\frac{\sum_{j,l}\alpha_j\alpha_l^*\langle b|a_j\rangle\langle a_l|b\rangle g\left(\frac{(a_j +a_l)}{2}-\frac{i \Omega(a_j-a_l)}{2}\right)\exp \bigl( \frac{-(a_j-a_l)^2\Delta_{\Pd }^2 g^2}{2}+i g(a_j-a_l)\mu_{\Pd }\bigr) }{\sum_{j,l}\alpha_j\alpha_l^*\langle b|a_j\rangle\langle a_l|b\rangle \exp\bigl( \frac{-(a_j-a_l)^2\Delta_{\Pd }^2 g^2}{2}+i g(a_j-a_l)\mu_{\Pd }\bigr) } \,.
\end{align}
\end{widetext}

The mean momentum of the device after the weak measurement and postselection on $b$ is 
\begin{equation}
\label{eq:16m}
\begin{split}
\langle \hat{\Pd}\rangle_b =\frac{\sum_{j,l}\alpha_j\alpha_l^{*}\langle b|a_j\rangle\langle a_l|b\rangle \int \Pd \tilde{\phi}(\Pd ,a_j)\tilde{\phi}^*(\Pd ,a_l) d\Pd }{\sum_{j,l}\alpha_j\alpha_l^*\langle b|a_j\rangle\langle a_l|b\rangle\int\tilde \phi(\Pd , a_j)\tilde\phi^*(\Pd , a_l) d\Pd }\\
=\frac{\sum_{j,l}-i g\Delta_{\Pd }^2 \alpha_j\alpha_l^*\langle b|a_j\rangle\langle a_l|b\rangle (a_j -a_l) e^{\frac{-(a_j-a_l)^2\Delta_{\Pd }^2 g^2}{2}}}{\sum_{j,l}\alpha_j\alpha_l^*\langle b|a_j\rangle\langle a_l|b\rangle e^{\frac{-(a_j-a_l)^2\Delta_{\Pd }^2 g^2}{2}}}\\
\end{split}
\end{equation}
where  $\tilde{\phi}(\Pd,a_i)=\exp(\Pd^2\Delta_{\Pd}^2(1+i\Omega)/4 -i \Pd g a_i)$ is the Fourier transform of $\phi(\Qd-ga_i)$ up to a normalisation constant.

\subsubsection{Small uncertainty $\Delta_P$}
\label{sec:lim1}

We now consider these expressions using the first method to obtain weak measurements, wherein the initial position of the measurement device becomes highly uncertain.  In the limit of $\Delta_P\to0$, we also have $\Omega\to0$. Consider the case where the mean momentum of the device, $\mu_{P}$, is also set to zero.  Using Eq.~(\ref{eq:14l}), the mean position of the device is
\begin{align}
\label{eq:15mml1}
\langle\hat{\Qd}\rangle_b &=\frac{\sum_{j,l}\alpha_j\alpha_l^*\langle b|a_j\rangle\langle a_l|b\rangle g\frac{(a_j +a_l)}{2}}{\sum_{j,l}\alpha_j\alpha_l^*\langle b|a_j\rangle\langle a_l|b\rangle} \nonumber \\
&=g \text{Re}[\langle\hat{A}_W\rangle] + g \Omega \text{Im}[\langle\hat{A}_W\rangle] \,,
\end{align}
where we have ignored terms of order $g^3 \Delta_P^2$ and higher as a result of taking $\Delta_P$ to be small.  In the limit $\Delta_P \to 0$, the covariance $\Omega \to 0$ and this shift becomes
\begin{equation}
\lim_{\Delta_P \to 0} \langle\hat{\Qd}\rangle_b =  g \text{Re}[\langle\hat{A}_W\rangle]\,.	
\end{equation}
From Eq.~(\ref{eq:16m}), the mean momentum of the device is
 \begin{align}
\langle\hat{\Pd}\rangle_b &=\frac{\sum_{j,l}-i g\Delta_{\Pd}^2\alpha_j\alpha_l^*\langle b|a_j\rangle\langle a_l|b\rangle (a_j -a_l)}{\sum_{j,l}\alpha_j\alpha_l^*\langle b|a_j\rangle\langle a_l|b\rangle} \nonumber \\
&=2 g \Delta_{\Pd}^2 \text{Im}[\langle\hat{A}_W\rangle]
\label{eq:17nl1}
\end{align}
which in the limit becomes
\begin{equation}
\lim_{\Delta_P \to 0} \langle\hat{\Pd}\rangle_b =  0 \,.	
\end{equation}

In this limit, the momentum of the device remains unchanged, and there is no disturbance to the state of the particle.  There is, however, a shift in the mean position of the device proportional to the real part of the weak value $\langle\hat{A}_W\rangle$. Because this limit implies $\Delta_{Q}\to\infty$, it is a shift in a uniform distribution and hence not physically resolvable.  
 
\subsubsection{Weak coupling $g$}
 
Consider now the second method for obtaining weak measurements, where the coupling strength $g$ is small and  the mean momentum of the device, $\mu_{P}$, is also set to zero. In the limit of $g\to0$, there is no disturbance to the system, however, in this limit, there is also no shift in the average position and momentum of the post selected devices. Hence, we then calculate the mean position and momentum of the measurement device after postselection to leading order in $g$.  The mean position of the device after postselection is
\begin{align}
\label{eq:15mm}
\langle\hat{\Qd}\rangle_b &=\frac{\sum_{j,l}\alpha_j\alpha_l^*\langle b|a_j\rangle\langle a_l|b\rangle g[\frac{(a_j +a_l)}{2}-\frac{i \Omega(a_j-a_l)}{2}]}{\sum_{j,l}\alpha_j\alpha_l^*\langle b|a_j\rangle\langle a_l|b\rangle} \nonumber \\
&=g \text{Re}[\langle\hat{A}_W\rangle]+ g \Omega \text{Im}[\langle\hat{A}_W\rangle] \,,
\end{align} 
and the mean momentum of the device after postselection is
\begin{align}
\label{eq:17n}
\langle\hat{\Pd}\rangle_b &=\frac{\sum_{j,l}-i g\Delta_{\Pd}^2\alpha_j\alpha_l^*\langle b|a_j\rangle\langle a_l|b\rangle (a_j -a_l)}{\sum_{j,l}\alpha_j\alpha_l^*\langle b|a_j\rangle\langle a_l|b\rangle}\\
&=2 g \Delta_{\Pd}^2 \text{Im}[\langle\hat{A}_W\rangle].
\end{align}
(In Ref.~\cite{complex}, the expression for $\langle\hat{\Qd}\rangle_b$ is written as $\langle\hat{\Qd}\rangle_b =g \text{Re}[\hat{A}_W]+ g m\frac{d\Delta_{Q}}{dt} \text{Im}[\hat{A}_W]$, where $m$ is the mass of the measurement device. To relate the expression of Eq.~(\ref{eq:15mm}) with this one, observe that the covariance of the device can be written as mass times the change in variance of position of the device propagating under a free Hamiltonian. Note however that, at the time of interaction, the Hamiltonian is not that of a free particle.)

\subsubsection{Shifts in the postselected distributions of the measurement device}

Comparing the shift in the mean position $\langle\hat{\Qd}\rangle_b$ of the device in the two cases, we find a term that is proportional to the imaginary part of the weak value of variable $\hat{A}$.  This additional term depends on the covariance $\Omega$ of the device, and as we will see in Sec.~\ref{sec:level3b}, it arises due to postselection of the particles biasing the device distribution. There is also a shift in the mean momentum $\langle\hat{\Pd}\rangle_b$ of the device, proportional to the imaginary part of the weak value. Note that in a limit where the disturbance due to measurement goes to zero, such as, $g\to0$ or $\Delta_P\to 0$, this shift in the mean momentum disappears; and only in this limit of finite disturbance, do we see the manifestation of the imaginary part of the weak value. The reason for this will become clear in our classical analysis in section Sec.~\ref{sec:level3b}. 

\subsection{\label{sec:level2c} Weak value of gaussian states}

Here, we analyse the weak value in the special case where the particle is a one-dimensional canonical quantum system prepared in a state with a gaussian wavefunction.  It is this special case which will be directly comparable to the results presented in the next section, within a theory of classical mechanics with an epistemic restriction.  Let $\hat{\qs}$ and $\hat{\ps}$ be the position and momentum observables for the particle, satisfying $[\hat{\qs},\hat{\ps}]=i\hbar$.  Let the initial quantum state of the particle be
\begin{equation}
\label{eq:psi}
  |\Psi\rangle \propto \int \exp\Bigl(-\frac{(\qs-\mu_{\qs})^2}{4\sigma^2}+i \mu_{\ps} \qs\Bigr) |\qs \rangle d\qs \,.
\end{equation}
This is a gaussian wavefunction with position mean $\mu_{\qs}$ and variance $\sigma^2$, and momentum mean $\mu_{\ps}$ and variance $\frac{1}{4\sigma^2}$.  While this state has been chosen to have zero convariance, we emphasise that our results are completely general (see note below).  The particle and the measuring device are coupled under the Hamiltonian in Eq.~\eqref{eq:4d}, where the observable $\hat{A}$ we are measuring is one of the form 
\begin{equation}
  \hat{A}=\cos\theta_A \hat{\qs}+\sin\theta_A \hat{\ps}.
\end{equation}

We then postselect using a projective measurement of the observable $\hat{B} = \cos{\theta_B}\hat{\qs} +\sin{\theta_B}\hat{\ps}$, with eigenstate
\begin{equation}
\langle b|=\int \sqrt{\frac{1}{2\pi i\sin\theta_B}} e^{-i(b \qs /\sin\theta_B - \cot\theta_B \qs^2/2)}\langle\qs| d\qs \,.
\end{equation}
The weak value of $\hat{A}$ for this postselection is
\begin{align}
\langle\hat{A}\rangle_W &=\cos\theta_A\frac{ \langle b|\hat{\qs}|\Psi\rangle}{\langle b|\Psi\rangle}+\sin\theta_A\frac{ \langle b|\hat{\ps}|\Psi\rangle}{\langle b|\Psi\rangle} \nonumber \\
&=\frac{4 \sigma^4 \cos\theta_B(b \cos\theta_A - \mu_{\ps } \sin(\Delta\theta))}{4 \sigma^4 \cos^2\theta_B+ \sin^2\theta_B} \nonumber \\  
&\quad+\frac{\sin\theta_B (\mu_{\qs } \sin(\Delta\theta) + b \sin\theta_A)}{4 \sigma^4 \cos^2\theta_B+ \sin^2\theta_B} \label{eq:final1} \\
    &\quad+ i \frac{2 \sigma^2 (-b + \mu_{\qs } \cos\theta_B + \mu_{\ps } \sin\theta_B) \sin(\Delta\theta)}{
4 \sigma^4 \cos^2\theta_B + \sin^2\theta_B}, \nonumber
\end{align}
where $\Delta\theta = \theta_B -\theta_A$.  We note that this expression depends linearly on $\mu_{\qs}$, $\mu_{\ps}$, and $b$, and while the real part of this expression has the same form as expected from Bayes' rule, the imaginary components of $\langle\hat{A}\rangle_W$ is also clearly identified.

\textit{Note:}  In Eq.~(\ref{eq:psi}), we consider a gaussian state with zero covariance for simplicity. However, this choice is equivalent to using a gaussian state with nonzero covariance simply by changing the quadratures of of both weak and strong measurement appropriately, that is, changing $\theta_A$ and $\theta_B$. Hence our analysis holds for general gaussian states.

\section{Weak values in the ERL theory}
\label{sec:level3}

In this section, we will analyse the weak measurement and postselection procedure described above in the context of a theory with a clear classical ontology:  the epistemically restricted Liouville (ERL) theory of Ref.~\cite{erl}.  This theory describes particles evolving in a phase space according to classical equations of motion.  What makes this theory interesting is an epistemic restriction that limits the knowledge that an observer can possess about the state of these particles.  It has been shown that there is a complete operational equivalence between the dynamics of the restricted Liouville distribution that describes an observer's knowledge in the ERL theory, and that of a subset of quantum theory, namely, gaussian quantum mechanics.  

For a full description of the ERL theory, the form of the epistemic restriction, and its consequences, see Ref.~\cite{erl}.  Briefly, the classical state of the particles in the theory are points in a phase space, i.e., positions and momenta.  An observer's knowledge about the state of a particle is given by a Liouville distribution, i.e., a probability distribution on phase space.  These phase space distributions are mathematically equivalent to Wigner functions of gaussian quantum states and satisfy the uncertainty principle, in other words, gaussians whose covariance matrices satisfy the following relationship
\begin{equation}
  \label{eq:11i}
  {\gamma}+i\Sigma \geq 0,
\end{equation}
where $\gamma$ is the covariance matrix defined as
\begin{equation}
  {\gamma}= \begin{bmatrix} 
  2\Delta_{q}^2 & \langle p q+q p\rangle-2\langle p\rangle \langle q\rangle\\  \langle p q+q p\rangle-2\langle p\rangle   \langle q\rangle & 2\Delta_{p}^2\\ \end{bmatrix}
\end{equation}
and $\Sigma$ is defined as
\begin{equation}
  \Sigma=\begin{bmatrix} 0 & 1\\ -1&0\\ \end{bmatrix}.
\end{equation}

In the remainder of this section, we will analyse weak measurements, postselection, and the corresponding weak value for gaussian states within ERL theory.

\subsection{Measurement model in ERL theory}
\label{sec:level3a} 

In this section, we will treat the measurement procedure outlined in section \ref{sec:level2a} within ERL theory. That is, we will interact the particles and measuring devices using a classical interaction and Hamilton's equations of motion. However, we will require that all initial Liouville distributions describing these systems obey the epistemic restriction~\eqref{eq:11i}. This restriction will lead us to a concept of weak measurement within ERL theory, including a tradeoff between information gain and disturbance much like quantum theory.  By following the structure of the quantum derivation in Sec.~\ref{sec:level2a}, we introduce a general measurement model within ERL theory and then investigate the situations under which the disturbance to the particles is minimised.

We model both the particle to be measured and the measurement device as one-dimensional canonical systems, with $\qs$ and $\ps$ the position and momentum coordinates of the particle, and $\Qd$ and $\Pd$ the position and momentum coordinates of the measurement device. The particle and the measurement device are coupled via the Hamiltonian
\begin{equation} 
  \label{eq:4dd} 
  H=\chi(t)\Pd A ,
\end{equation}
where the observable we are measuring is of the form
\begin{equation}
A=\cos\theta_A \qs +\sin\theta_A \ps \,.
\end{equation}
The distribution of the particle and the measurement device changes according the the classical Hamiltonian equations as a result of this interaction, following
\begin{equation}
 \frac{dq}{dt}=\frac{\partial H}{\partial p}\,, \quad \frac{dp}{dt}=-\frac{\partial H}{\partial q}\,.
\end{equation}
After the measurement, the position and momentum of the particle are
\begin{align}
  \label{eq:12a}
  \qs &=\qs_{i}+g \sin\theta_A \Pd_{i}\,, \\
  \label{eq:12b}
  \ps &= \ps_{i}-g \cos\theta_A \Pd_{i}\,,
\end{align}
where $\ps_{i}$ and $\qs_{i}$ are the initial momentum and position of the particle being measured respectively and $\Pd_{i}$ is the initial momentum of the measuring device.  

The position and momentum of the measurement device after this interaction are
\begin{align}
  \label{eq:12c}
  \Qd &=\Qd_{i}+g( \cos\theta_A \qs_{i}+ \sin\theta_A \ps_{i})\,, \\
  \label{eq:12d}
  \Pd &= \Pd_{i} \,.
\end{align}
Hence the change in the position of the device gives the measurement outcome. Note there is no change in the momentum of each device due to the measurement interaction, which suggests that any change to the mean momentum of the ensemble of devices, must be statistical bias.
 
For an ideal measurement, we would require the initial position of the device $\Qd_{i}$ to be known with complete certainty, and this would lead to a perfect correlation between the observable $A$ and the position of the measurement device after interaction. However, due to our epistemic restriction, the uncertainty in the momentum of the device must be infinite in this case, i.e., the observer has no knowledge of the initial momentum of the measurement device $\Pd_{i}$. From Eqs.~\eqref{eq:12a} and \eqref{eq:12b}, the position and momentum of the particle are each displaced by an amount proportional to the initial momentum of the device as a result of the interaction.  If the initial momentum $\Pd_i$ of the measurement device is unknown, so is the phase space displacement of the particle. Thus we see how, within the ERL theory, such a measurement that acquires complete information about an observable $A$ of the particle is accompanied by a corresponding unknown disturbance on the state of the particle. 

It is in this way that an information gain-disturbance tradeoff appears in the ERL theory, despite its classical ontology.  If we removed the epistemic restriction, the state of the particle would still be displaced by an amount proportional to the momentum of the device.  However, the observer could in principle know the initial value of the momentum of the device as well as its position and therefore correct for this change.  In other words, disturbance due to measurement arises in Newtonian mechanics, and the epistemic restriction just makes the disturbance unrectifiable.  Note that, despite the change in the position and momentum of the particles due to the interaction, the expectation value of the observable $A$ has not changed, i.e.,
\begin{align}
  A(\ps,\Pd) &= \cos\theta_A(\qs_{i} - g \sin\theta_A \Pd) +\sin\theta_A(\ps_{i}+g \cos\theta_A \Pd) \nonumber \\
  &= \cos\theta_A \qs_{i}+\sin\theta_A \ps_{i} \nonumber \\
  &=A(\ps_{i},\Pd_{i})\,.
\end{align}
As in quantum theory, measurements in ERL theory are repeatable, and the disturbance is associated with uncertainty in canonically conjugate observables.

\subsection{Weak measurements in ERL theory}

We now introduce weak measurements in the ERL theory, again following by analogy the quantum formalism of Sec.~\ref{sec:level2aa}.  From Eqs.~\eqref{eq:12a} and~\eqref{eq:12b}, one can again see two methods by which the disturbance due to measurement can be made small:  first, by considering small coupling $g$, and second, by requiring the initial momentum of the measurement device $\Pd_{i}$ to be very close to zero.  In the second method, to ensure that the momentum of the measurement device has $\Pd_{i}=0$ (and not just the mean value of the Liouville distribution), we must require that the variance $\Delta_{\Pd_i}$ is very small as well as the mean value $\mu_{\Pd_i}=0$.  Due to the epistemic restriction, the initial uncertainty in position of the measurement device $\Delta_{\Qd_{i}}$ must then be very large.  Therefore, in the limit of small $\Delta_{\Pd_i}$, we have vanishing knowledge of the initial position of the device and hence any change in this position after the measurement (Eq.~\eqref{eq:12c}).  Thus, in both methods, we obtain weak measurements that yield an arbitrarily small information gain about the system and correspondingly small disturbance. 

Within ERL mechanics, we note that these two methods of obtaining weak measurements are physically distinct.  In both cases, in the limits $g \to 0$ or $\Delta_P \to 0$ we have both no disturbance to the system as well as no information gained about the system.  There is a difference in the ontology, however.  In the limit of $g\to0$, there is no physical change to the measurement device.  In contrast, in the limit of $\Delta_{P}\to 0$, there is a shift in the mean position of the measurement device, but our uncertainty about the device's initial position makes the shift undetectable.

\subsection{The weak value in ERL theory}
\label{sec:level3b}

With a meaningful notion of weak measurement in ERL theory, we can now consider the appearance of a weak value.  Let the initial phase space distribution of the particles being measured be described by a position distribution with mean $\mu_{\qs }$ and variance $\sigma^2$ and a momentum distribution with mean $\mu_{\ps }$ and variance $\frac{1}{4\sigma^2}$, i.e., as the Liouville distribution
\begin{equation}
  \rho_s(\qs_{i} ,\ps_{i} )=\frac{1}{\pi} \exp\Bigl(\frac{-(\qs_{i} -\mu_{\qs })^2}{2\sigma^2}-(\ps_{i} -\mu_{\ps })^2 2 \sigma^2 \Bigr)\,.
\end{equation}
This distribution is precisely the Wigner function of the initial quantum state $|\Psi\rangle$ of Eq.~\eqref{eq:psi}.  The initial phase space distribution of the measurement device is the Wigner function of the initial state $|\Phi\rangle$ in Eq.~\eqref{eq:1a},
\begin{equation}
  \rho_d(\Qd_{i},\Pd_{i})=\frac{1}{\pi} \exp\Bigl(-(2\Delta_{\Pd}^2 \Qd_{i}^2 +2\Delta_{\Qd}^2 \Pd_{i}^2-2 \Omega \Pd_{i} \Qd_{i} )\Bigr),
\end{equation}
where $\Omega=2\langle \Pd_{i} \Qd_{i}\rangle-2\langle \Pd_{i}\rangle\langle \Qd_{i}\rangle$ form the off diagonal elements of the covariance matrix. 

Observable $A$ is measured using weak measurements as described above. After the weak measurement, the actual position and momentum of each particle and device changes according to Eqs.~\eqref{eq:12a}-\eqref{eq:12d}, and the phase space distributions of the particles and devices become correlated as
\begin{multline}
\label{eq:20a}
\rho'_{sd}(\qs,\ps,\Qd,\Pd) = \rho_s(\qs-g \sin\theta_A \Pd,\ps +g \cos\theta_A \Pd)\\
 \times\rho_d(\Qd-g( \cos\theta_A \qs +\sin\theta_A \ps),\Pd)\,.
\end{multline}

The particles are then measured using a strong measurement of observable $B = \cos{\theta_B}q +\sin{\theta_B}p$, and postselect on the outcome $\cos{\theta_B}q +\sin{\theta_B}p =b$. The postselected distribution of the measurement devices is then 
\begin{align}
\rho''_{d}(\Qd,\Pd)&=\int\int \rho'_{sd}(\qs,\ps,\Qd,\Pd)\delta(B=b) \,d\qs\, d\ps \\
&=\int \rho'_{sd}(\qs ,\frac{b-\cos\theta_B \qs }{\sin\theta_B},\Qd,\Pd) \,d\qs .
\end{align}
The mean position of the postselected subset is
%\begin{widetext}
\begin{align}
\label{eq:position}
\langle \Qd \rangle_b &=\int\int \Qd \rho''_{d}(\Qd,\Pd)\, d\Pd\, d\Qd \nonumber \\
&= \frac{\alpha(g,\theta_A,\theta_B,\Delta_{\Qd},\mu_{\ps},\sigma,\Omega,b)}{\beta(g,\theta_A,\theta_B,\Delta_{\Qd},\sigma,\Omega)} ,
\end{align}
where $\alpha$ and $\beta$ are real-valued functions defined as
\begin{align}
\alpha &(g,\theta_A,\theta_B,\Delta_{\Qd},\mu_{\ps},\sigma,\Omega,b) \nonumber \\ 
&=g^3(\mu_{\qs } \cos\theta_A + \mu_{\ps } \sin\theta_A)(1+\Omega^2)\sigma^2\sin^2(\Delta\theta) \nonumber  \\
&\quad +g\Delta_{\Qd }^2 \Bigl(4\sigma^4 \cos\theta_B(b \cos\theta_A - \mu_{\ps } \sin(\Delta\theta)) \nonumber \\
&\qquad +\sin\theta_B (\mu_{\qs } \sin(\Delta\theta)+b \sin\theta_A) \nonumber \\
&\qquad +2\Omega \sigma^2 \sin(\Delta\theta)(\mu_{\qs } \cos\theta_B+ \mu_{\ps } \sin\theta_B - b) \Bigr) \\
\beta &(g,\theta_A,\theta_B,\Delta_{\Qd},\sigma,\Omega) \nonumber \\
&= g^2(1+\Omega^2)\sigma^2\sin^2(\Delta\theta)+\Delta_{\Qd }^2(4\sigma^4\cos^2\theta_B+\sin^2\theta_B) .
\end{align}
%\end{widetext}
The mean momentum of the postselected subset is
%\begin{widetext}
\begin{align}
\label{eq:momentum}
\langle \Pd \rangle_b &=\int\int \Pd \rho''_{d}(\Qd,\Pd)\, d\Pd\, d\Qd \nonumber \\
&=\frac{g (1+\Omega^2)\sigma ^2( \mu_{\qs } \cos\theta_A+\mu_{\ps } \sin\theta_A-b) \sin(\Delta\theta)}{\beta(g,\theta_A,\theta_B,\Delta_{\Qd},\sigma,\Omega)}.
\end{align}
%\end{widetext} 

\subsubsection{Small uncertainty $\Delta_P$}

As with the quantum case, we consider these expressions using the first method to obtain weak measurements, wherein the initial position of the measurement device becomes highly uncertain.  We characterise this case by choosing $\mu_p = 0$ and take the limiting case $\Delta_P\to 0$, which implies $\Omega\to0$ as well.  We find we can express the average shift in the position of the measurement devices upon postselection using the same expression for $\langle \hat{A}\rangle_W$ as calculated in Eq.~\eqref{eq:final1}, to give 
\begin{equation}
  \langle \Qd \rangle_b = g \text{Re} [\langle \hat{A}\rangle_W]+ g \Omega \text{Im}[\langle\hat{A}_W\rangle] \,,
\end{equation}
where again we have ignored terms of order $g^3 \Delta_P^2$ and higher.  In the limit $\Delta_P \to 0$, the covariance $\Omega \to 0$ and this becomes
\begin{equation}
\lim_{\Delta_P \to 0} \langle\hat{\Qd}\rangle_b =  g \text{Re}[\langle\hat{A}_W\rangle]\,.	
\end{equation}
This exactly reproduces the quantum mechanical shift in mean position. As our model has a classical ontology, it allows joint probability distributions over all variables, unlike quantum mechanics. As a result, we are able to calculate conditional expectation values of any two observables.  In our situation, we can exploit this fact to compare the shift in the average position of the device with the expectation value of $A$ of the particles conditioned on an outcome $b$ of observable $B$.  We find that the real part of the weak value is indeed the same as the conditional expectation value. 

In this limit we find that the mean momentum of the devices is
\begin{equation}
   \langle\Pd\rangle_b = 2 g \Delta_{\Pd}^2 \text{Im}[\langle\hat{A}_W\rangle] \,,
\end{equation}
which in the limit becomes
\begin{equation}
\lim_{\Delta_P \to 0} \langle\hat{\Pd}\rangle_b =  0 \,.	
\end{equation}
Again, this exactly reproduces the quantum mechanical expression.

\begin{figure}
\includegraphics[width=0.45\textwidth]{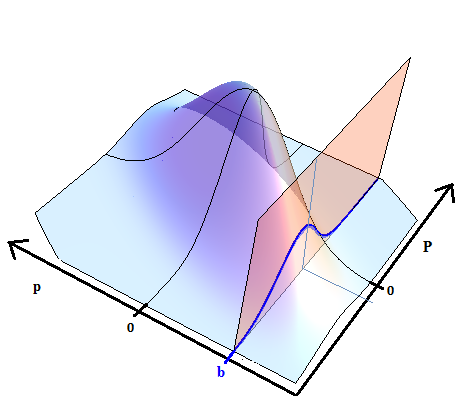}
\caption{Joint probability distribution of the momenta of the particle and measurement device after weak measurement.  The position $p$ of the particles are weakly measured with devices sampled from a momentum distribution with mean zero and $\Delta_P \neq 0$. Each particle's momentum is shifted by an amount proportional to the momentum of the device $P$, following~\eqref{eq:12b}. If one now postselects measurement devices based on the particle having momentum $p=b$, then the distribution of these devices is shown on the plane intersecting the plot. The mean of this postselected distribution is no longer zero.
}\label{fig:fig2}
\end{figure}

\subsubsection{Weak coupling $g$}

Just as in the quantum case, we take the mean momentum $\mu_P = 0$.  If we take $g$ to be finite but small enough to ignore higher than first order terms of $g$, the shift in the average position of the postselected devices is
\begin{equation}
  \langle \Qd \rangle_b = g \text{Re} [\langle \hat{A}\rangle_W] +g \Omega \text{Im} [\langle \hat{A} \rangle_W ]\,,
\end{equation}
and the average momentum shift is
\begin{equation}
  \langle \Pd \rangle_b = 2 g\Delta_{\Pd}^2 \text{Im}[\langle\hat{A}\rangle_W].
\end{equation}
From Eq.~\eqref{eq:12d}, we see that the momentum of the individual measurement devices are not changed due to the weak measurement. However in this method for weak measurements, we do not require $\Delta_P\to0$, i.e., there is uncertainty of the initial momentum of the measurement devices. The state of the particles can be shifted in phase space by the momentum of the devices, as shown in \eqref{eq:12a} and \eqref{eq:12b}, and so initial uncertainty in the momentum of the measurement device leads to an unknown disturbance of the particles. By postselecting on particles that have been perturbed by the momentum of the device, we arrive at a final momentum distribution of the device that is biased. 

An example of this effect is illustrated in Fig.~2, where the position of the particle is weakly measured, followed by postselection on the particle's momentum, that is, $\theta_{A}=0$ and $\theta_{B}=\frac{\pi}{2}$.  The shift in the mean momentum of the measurement devices does not arise as a result of the dynamics of the interaction, but instead from postselection biasing the momentum distribution.  If the position distribution of the devices is correlated to the momentum distribution via a nonzero covariance $\Omega$, the position distribution of the devices will also be biased. 

In summary, we have seen how terms in both the position and momentum shifts that are proportional to the imaginary part of the weak value are primarily the result of postselection biasing the device distributions, not dynamics.  This observation allows us to formulate an operational interpretation of the imaginary part of the weak value as a measure of how much postselection will bias the device distribution, given finite disturbance, which we emphasise results from uncertainty in the initial properties of the devices.

\subsection{Bounds on the weak measurement regime} 
\label{sec:level5}

As an additional insight that arises from our analysis of the weak value, we now provide a better bound on the range of couplings $g$ for which the standard weak value results will hold.  In order to consider $g$ to be small (and therefore to ignore higher order terms), from equations \eqref{eq:14l}, \eqref{eq:15mm}, \eqref{eq:16m}, \eqref{eq:17n}, one can see that this approximation is good provided
\begin{equation}
  \label{limit1}
  g^2\Delta_{\Pd }^2 \ll\frac{1}{  (a_j -a_l)^2 }\quad \forall j,l,
\end{equation}
  where $a_i$ is an eigenvalue of the measurement operator.  This condition is very restrictive and is not useful for variables with a continuous eigenspectrum.  However, we can derive a less restrictive upper bound by looking at the approximations made on the classical distributions of the devices. From Eqs.~\eqref{eq:position} and\eqref{eq:momentum}, we can see that in order to ignore higher order terms of $g$, we require
\begin{equation}
  g^2\Delta_{\Pd }^2\ll\frac{(4\sigma^4 \cos^2\theta_B +\sin^2\theta_B)}{4\sigma^2\sin^2(\theta_A-\theta_B)}\,.
\end{equation}
Because ERL theory is operationally equivalent to gaussian quantum mechanics, this upper bound is also true for gaussian quantum states, and could equally well be derived from the quantum formalism. This is a significantly less restrictive upper bound for gaussian states than the one discussed in Ref.~\cite{duck}.

\section{Conclusion}
\label{sec:level6}

The weak value has long been argued to be a fundamentally quantum phenomenon.  Here, we have analysed the weak value in a theory with a clear classical ontology but one in which information about a (classical system) is limited.  This epistemically-restricted theory provides an analogy of the weak value:  one which is exact when compared with weak value experiments in gaussian quantum mechanics.  Within this epistemically-restricted theory, we see the same average shifts in the position and momentum of the devices as in our quantum analysis.  Because our ERL model has a clear ontology, it gives us insight into the statistical effects of postselecting on weakly disturbed states. We find that the real part of weak value is the conditional expectation and the imaginary part of the weak value is a measure of how much post selection biases the distribution of the system being measured given any finite disturbance.

We do not see the appearance of any anomalous weak values in our analysis. This is because in addition to the observables in our model having an unbounded spectrum, our analysis is restricted to gaussian quantum mechanics, which is known to be non-contextual as it allows a non-negative quasiprobability representation (the Wigner function).   Our interpretation of the real and imaginary parts of the weak value cannot be naively extended to states and measurements that allow for the observation of anomalous weak values, as these would imply a proof of contextuality that explicitly rules out the existence of the type of epistemically-restricted classical mechanics we employ.  What our results show is that for states and measurements that are noncontextual, the weak measurement procedure followed by post selection reproduces shifts proportional to the real and imaginary parts of the weak value even in a model based on classical mechanics.  Our work also suggests that quasiprobabilistic representations might prove to be a useful tool in analysing the weak value for the more general case.

\begin{acknowledgments}
We thank Chris Ferrie, Nick Menicucci, Rafael Alexander and Harrison Ball for interesting discussions and helpful feedback.  This work is supported by the ARC via the Centre of Excellence in Engineered Quantum Systems (EQuS) project number CE110001013.
\end{acknowledgments}

\end{document}